\documentclass[lineno]{biometrika}

\usepackage{amsmath}
\usepackage{amssymb}
\usepackage{mathabx}

\usepackage{times}
\usepackage{bm}
\usepackage{natbib}
\usepackage{tikz}
\usepackage{pgf}
\usepackage{pgfplots}
\nolinenumbers

\usetikzlibrary{patterns}
\usetikzlibrary{shapes}
\usetikzlibrary{arrows,decorations.pathmorphing,backgrounds,fit,positioning,shapes.symbols,chains}
\usetikzlibrary{shapes.geometric}

\definecolor{MyDarkBlue}{rgb}{0.1,0,0.40}
\definecolor{MyLightBlue}{rgb}{0, 229, 238}
\definecolor{MyMaize}{rgb}{255, 255, 0}
\definecolor{MyLemon}{rgb}{250, 250, 205}
\definecolor{MyDarkRed}{rgb}{178, 48, 96}

\tikzstyle{qShape} =  [draw=black, very thick, align=left, diamond,
minimum width=4em]
\tikzstyle{format} = [draw, thin, fill=blue!20]
\tikzstyle{medium} = [ellipse, draw, thin, fill=green!20, minimum height=2.5em]
\tikzstyle{rand} = [circle, draw=red, very thick, minimum height=2.5em]
\tikzstyle{cond} = [draw, thin]
\tikzstyle{mybox} = [draw=MyDarkBlue, very thick, align=left,
    rectangle, rounded corners, inner sep=3pt, inner ysep=10pt]
\tikzstyle{qbox} = [draw=MyDarkBlue, very thick, fill=blue!5,
    rectangle, rounded corners, inner sep=15pt, inner ysep=35pt]

\tikzstyle{sbox} = [draw=MyDarkBlue, very thick, fill=blue!5,
    rectangle, rounded corners, inner sep=1pt, inner ysep=12pt]

\tikzstyle{fancytitle} =[fill=MyDarkBlue, text=white]
\tikzstyle{taylor} = [rectangle, draw=MyDarkBlue, thick,
    minimum height=2.5em, inner sep=2pt, inner ysep=1pt]
\tikzstyle{plain} = [draw=none, fill=none]

\usepackage[plain,noend]{algorithm2e}

\newcommand{\OMIT}[1]{\relax}   
\def\text{{\rm}}

\def\t{T}

\newcommand{\rp}{\mathbb{R}^p}

    \newcommand\independent{\protect\mathpalette{\protect\independenT}{\perp}}
    \def\independenT#1#2{\mathrel{\rlap{$#1#2$}\mkern2mu{#1#2}}}

\makeatletter
\renewcommand{\algocf@captiontext}[2]{#1\algocf@typo. \AlCapFnt{}#2} 
\def\@algocf@capt@plain{top}
\renewcommand{\algocf@makecaption}[2]{%
  \addtolength{\hsize}{\algomargin}%
  \sbox\@tempboxa{\algocf@captiontext{#1}{#2}}%
  \ifdim\wd\@tempboxa >\hsize
    \hskip .5\algomargin%
    \parbox[t]{\hsize}{\algocf@captiontext{#1}{#2}}
  \else%
    \global\@minipagefalse%
    \hbox to\hsize{\box\@tempboxa}
  \fi%
  \addtolength{\hsize}{-\algomargin}%
}
\makeatother

\def\Bka{{\it Biometrika}}

\def\T{{ \mathrm{\scriptscriptstyle T} }}

\begin{document}

\markboth{Hu et al.}{Regret bounds for Thompson sampling}

\title{Note on Thompson sampling for large decision problems}

\author{Tao Hu, Eric B. Laber, Zhen Li, Nick J. Meyer}
\affil{Department of Statistics, North Carolina State University, Raleigh, North Carolina 27606, U.S.A. \email{thu3@ncsu.edu} \email{laber@stat.ncsu.edu}}
\author{and Krishna Pacifici}
\affil{Department of Applied Ecology, North Carolina State University, Raleigh, North Carolina 27606, U.S.A. \email{jkpacifi@ncsu.edu} }

\maketitle

\begin{abstract}
There is increasing interest in using streaming data to inform 
decision making across a wide range of application domains including
mobile health, food safety, security, and resource management.   A
decision support system formalizes online decision making as a map
from up-to-date information to a recommended decision.   Online estimation
of an optimal decision strategy from streaming data requires
simultaneous estimation of components of the underlying system
dynamics as well as the optimal decision strategy given these dynamics; thus, 
there is an inherent trade-off between choosing decisions that lead to 
improved estimates and  choosing decisions that appear to be
optimal based on current estimates.   Thompson (1933) was 
among the first to formalize this trade-off in the context of choosing
between two treatments for a stream of patients; he proposed a simple
heuristic wherein a treatment is selected randomly at each time point with selection
probability proportional to the posterior probability that it is
optimal.  We consider a variant of Thompson sampling that is simple
to implement and can be
applied to large and complex decision problems.  We
show that the proposed Thompson sampling estimator is 
consistent for the optimal decision support system 
and provide rates of convergence and finite sample error bounds.  
The proposed algorithm is illustrated using an agent-based model 
of the spread of influenza on a network and management of mallard populations
in the United States.  
\end{abstract}   


\section{Introduction}
Technological advancements have made it possible to collect, store,
manipulate, and access large amounts of data on complex systems in
real-time.  Consequently, there is enormous potential to use
accumulating data to construct adaptive decision support systems that
map up-to-date information to a recommended decision.  For example, in
the context of mobile-health, data collected both passively and
actively through a mobile device can be used to monitor a patient's
health status and to construct an individualized treatment strategy
that applies interventions if, when, and in the amount they are needed
\citep[][]{riley2011health, litvin2013computer, kumar2013mobile,
  spruijt2014dynamic, nahum2014just}.  Other examples include
data-driven management of infectious diseases wherein accruing
information about the spread of the disease can be used to inform how
best to allocate treatment resources \citep[][]{chades2011general,
  meyer2016}, and adaptive management of natural resources wherein
management decisions are adjusted over time according to current and
forecasted resource availability \citep[][]{mccarthy2010resource,
  mcdonald2011allocating, marescot2013complex, fackler2014addressing}.

To estimate a decision support system that maximizes mean cumulative
utility, we apply a variant of Thompson sampling
\citep[][]{thompson1933likelihood} that avoids directly computing a
posterior distribution over the optimal decision at each time point.
This estimator is computationally efficient and can be applied in
settings in which data are: (i) accumulating rapidly over an
indefinite time horizon; (ii) high-dimensional; (iii) composed of a
single data stream, i.e., no independent replication; and (iv) the
number of possible decisions is too large to enumerate.  
We derive rates of convergence on the difference
in cumulative utility under the proposed estimator and an optimal
decision support system.
The proposed
estimator relies on a model for the underlying system dynamics and
therefore is ideally suited to settings where existing domain
knowledge or historical data can be used to inform a class of
models. In our motivating applications, such domain knowledge is
abundant.   
 In settings where domain knowledge is scarce, 
 the proposed methodology can be extended 
 to accomodate more flexible models that grow in complexity
 as data accumulate.

The estimation problem we consider here is related to estimation of an
optimal dynamic treatment regime \citep[][]{murphy2003optimal,
  robins2004optimal, chakraborty2013statistical, kosorok2015adaptive}.
Like the decision support systems we consider, a dynamic
treatment regime is a sequence of functions, one per decision stage,
that map up-to-date information to a recommended decision and the goal
is to estimate a regime that maximizes expected cumulative utility.
However, existing methodology for estimation of optimal dynamic
treatment regimes is designed for application to data collected in
observational or randomized studies involving a cohort of patients.
Thus, almost all methodology for dynamic treatment regimes is designed
for offline estimation using data composed of independent, identically
distributed replicates of the decision process observed over a finite
time horizon.  In contrast, the problems we consider here involve
online estimation using a single stream of data, and an indefinite
time horizon.  Some methodology for dynamic treatment regimes touches
on at least one of these features: \citet{ertefaie2014constructing}
proposed a variant of the $Q$-learning algorithm
\citep[][]{murphy2005generalization, schulte2014q} that applies to
problems with an indefinite time horizon but this methodology is
designed for offline estimation using a batch of independent,
identically distributed replicates; \citet{actorCritic} proposed a
policy-search algorithm \citep[][]{zhang2012robust,
  zhao2012estimating, zhao2015new} that applies to indefinite time
horizons but requires independent, identically distributed replicates;
and \citet{minsker2015active} proposed to use an online estimator of
an optimal treatment regime to adaptively change patient recruitment
probabilities, however this also requires replicates and only applies
to a single decision point.

The proposed estimator is an example of a model-based planning algorithm
in reinforcement learning \citep[][]{sutton1998reinforcement,
  powell2007approximate}.  Model-based planners estimate a system
dynamics model and then apply (approximate) dynamic programming
algorithms to the estimated system as if it were known.  A key feature
of model-based planning is the need to balance making decisions that
lead to improved model estimates with those that lead to high-utility
under the current estimated model; in the computer science literature this
is known as the exploration-exploitation trade-off
\citep[][]{kaelbling1996reinforcement, sutton1998reinforcement}.
Thompson sampling has been studied extensively as a means of balancing
exploration and exploitation in the context of multi-armed bandit
problems \citep[][]{agrawal2012analysis,kaufmann2012thompson,
  korda2013thompson, agrawal2013thompson, gopalan2014thompson,
  russo2014information}.  However, Thompson sampling for more complex
decision problems in which the decisions affect not only immediate
utility but also the state of the system and subsequently future
potential for utility, has received considerably less attention.
\citet{gopalan2015thompson} applied Thompson sampling to Markov
decision processes and derived convergence rates similar to those
presented here.  However, the variant of Thompson sampling proposed by
\citet{gopalan2015thompson} has several features that prevent direct
application to our setting; their algorithm requires
that: (i) the set of system states be finite and that the underlying
decision process returns infinitely often to a fixed reference state,
in the settings we consider, e.g., control of an infectious
disease, the state is continous and there is no guarantee of return to
a reference state; (ii) a fixed policy be applied for prolonged
periods in which the estimated system dynamics model is improved, this
may not be feasible or ethical in settings with human subjects or
limited natural resources; and (iii) one be able to efficiently
compute draws from the posterior which may not be possible without
conjugate priors.  Analyses of the operating characteristics of
Thompson sampling from a Bayesian perspective are given in
\citet[][]{osband2013more} and \citet{osband2014model}.

In Section 2, we introduce an approximate Thompson sampling algorithm
for parametric models.  In Section 3, we provide rates of convergence
for this variant of the Thompson sampling algorithm.  In Section 4, we
illustrate the use of Thompson sampling using a simple agent-based model of
influenza and a model for management of mallard populations in the
United States.  Section 5 contains a discussion of open problems and
concluding remarks.

\section{Thompson sampling with parametric models}
We consider a decision problem evolving in discrete time
$\mathcal{T}= \left\lbrace 1, 2,\ldots\right\rbrace$.  At each time
$t\in\mathcal{T}$, the decision maker: (i) observes the current state
of the process $S^t\in\rp$; (ii) selects an action
$A^t\in\mathcal{A}$; and (iii) observes next state $S^{t+1}$ and
utility $U^t = U(S^{t+1}, A^t, S^t) \in\mathbb{R}$.  We assume (A0)
that the state process is Markovian so that $S^t$ is conditionally
independent of $S^1,\ldots, S^{t-2}, A^1,\ldots, A^{t-2}$ given
$S^{t-1}$ and $A^{t-1}$; in some settings, for this definition to hold
the state $S^t$ might contain features constructed from observations
and actions collected over multiple time points not just information
collected between the $(t-1)$st and $t$th decision.  A decision
strategy, $\pi:\mathrm{dom}\,S^t\rightarrow \mathrm{dom}\,A^t$, is a
map from states to actions so that under $\pi$ a decision maker
presented with $S^t=s^t$ at time $t$ will selection action
$\pi(s^t)$. An optimal decision strategy maximizes mean discounted
utility if applied to select actions in the population of interest;
a formal definition is given below.
Our goal is to construct
an estimator of an optimal strategy that can be applied when the
dimension of the state space, $p$, and the number of possible actions,
$\mathrm{card} \left(\mathcal{A}\right)$, are large.  For example, in
spatial-temporal applications, like the influenza example presented in
Section 4, $p$ may be on the order of tens of thousands and
$\mathrm{card}\left(\mathcal{A}\right)$ is exponential in $p$.

We use potential outcomes \citep[][]{rubin1974estimating} to define
an optimal decision strategy. We use an overline to denote history, i.e.,
$\overline{a}^{t} = (a^1,\ldots, a^t)$.  The potential state
at time $t$ under action sequence $\overline{a}^{t-1}$ is  denoted
$S^{*t}(\overline{a}^{t-1})$; thus, the potential utility at time
$t$ is $U^{*t}(\overline{a}^t) = 
U\left\lbrace S^{*t+1}(\overline{a}^{t}), {a}^t,
S^{*t}(\overline{a}^{t-1})
\right\rbrace$.  For any strategy, $\pi$, the potential state at time
$t$ under $\pi$ is $S^{*t}(\pi) = 
\sum_{\overline{a}^{t-1}}S^{*t}(\overline{a}^{t-1})\prod_{v=1}^{t-1}
1_{\pi\left\lbrace S^{*v}(\overline{a}^{v-1})\right\rbrace = a^v}$ and subsequently
the potential utility is $U^{*t}(\pi) = U\left[
S^{*t+1}(\pi), \pi\left\lbrace S^{*t}(\pi)\right\rbrace, S^{*t}(\pi)
\right]$.  Define the total discounted
mean utility of a strategy $\pi$ as
\begin{equation}\label{valueFn}
V(\pi) = \mathbb{E}\sum_{t\ge 1}\gamma^{t-1}U^{*t}(\pi),
\end{equation}
where $\gamma \in (0,1)$ is a discount factor that balances proximal
and distal utility \citep[][]{sutton1998reinforcement,
  puterman2014markov}.  Given a class of strategies, $\Pi$, an
optimal strategy, $\pi^{\mathrm{opt}}\in\Pi$, satisfies
$V(\pi^{\mathrm{opt}}) \ge V(\pi)$ for all $\pi \in \Pi$. Thus, an
optimal regime is defined in terms of the class $\Pi$ which may be chosen
to enforce parsimony, logistical or cost constraints, or other
structure.  Hereafter, we consider $\Pi$ as fixed and known; in
practice, the choice of an appropriate class of strategies will depend
on the domain of application, see Section 4 for examples.

To ensure that $\pi^{\mathrm{opt}}$ is identifiable in terms of the
underlying generative model, we make a series of standard assumptions
\citep[][]{robins2004optimal, schulte2014q}.  Let
$W^* = \left\lbrace S^{*t}(\overline{a}^{t-1}),
  U^{*t}(\overline{a}^t)\,:\,
  \overline{a}^{t}\in\mathcal{A}^{t}\right\rbrace_{t\in\mathcal{T}}$
denote the set of all potential states and utilities.  We assume: (A1)
sequential ignorability,
$W^*\independent A^t \mid \overline{S}^t, \overline{A}^{t-1}
\,\,(t \in \mathcal{T})$; (A2)
positivity,
$P\left(A^t = a^t | \overline{S}^t= \overline{s}^t,
  \overline{A}^{t-1}=\overline{a}^{t-1}\right) > 0$
for all $\overline{s}^t$, $\overline{a}^{t-1}$ such that
$a^t \in \left\lbrace
  \pi(s^t)\,:\,\pi\in\Pi\right\rbrace\,\,(t\in\mathcal{T})$;
and (A3) consistency,
$S^t = S^{*t}(\overline{A}^{t-1})\,\,(t \in \mathcal{T})$.   Under
(A0)-(A3)  it can be seen that the mean utility at time $t$ under $\pi$ is
\begin{equation}\label{integralEqn}
\mathbb{E}U^{*t}(\pi) = 
\int U(s^t, a^{t-1}, s^{t-1}) 
\left\lbrace
\prod_{v=2}^{t}f^t(s^v|s^{v-1},
a^{v-1})\delta_{\pi(s^{v})}(a^v)
\right\rbrace \delta_{\pi(s^1)}(a^1)f^1(s^1)
d\lambda(\overline{s}^t, \overline{a}^t),
\end{equation}
where $f^v$ is the conditional density of $S^v$ given $S^{v-1}$ and
$A^{v-1}\,(v=2,\ldots, t)$, $f^1$ is the marginal density of $S^1$,
$\delta_{u}$ is a point mass at $u$, and $\lambda$ is a dominating
measure.  The right-hand side of (\ref{integralEqn}) is a
functional of the underlying generative model and given estimators of
the densities $f^v$ for $v=2,\ldots, t$ one could construct a plug-in
estimator of $\mathbb{E}U^{*t-1}(\pi)$.  However, non-parametric
estimation of these densities is not possible in general as there is
only a single observation per time point.  Thus, to facilitate
estimation, we impose further structure on these densities.

We assume that the densities $f^t$ for $t \ge 2$ are stationary and indexed
by a low-dimensional parameter
$\theta \in \Theta \subseteq \mathbb{R}^q$, i.e.,
$f^t(s^t|s^{t-1}, a^{t-1}) = f(s^{t}|s^{t-1}, a^{t-1}; \theta^*)$,
where $f$ is not indexed by $t$ and $\theta^*\in\Theta$ denotes the
true parameter value.  The likelihood for $\theta$ is
\begin{eqnarray*}
\mathcal{L}_{t}(\theta) &=& 
\left\lbrace
\prod_{v=2}^{t}f(S^v\mid S^{v-1}, A^{v-1}; \theta)
p^{v-1}\left(A^{v-1}|\overline{S}^{v-1}, \overline{A}^{v-2}\right)
\right\rbrace f^1\left(S^1\right) \\ &\propto & 
\prod_{v=2}^{t}f(S^v\mid S^{v-1}, A^{v-1}; \theta),
\end{eqnarray*}
where $p^v$ denotes the conditional distribution over actions used by
the decision maker and $p^{1}(A^1|\overline{S}^1, \overline{A}^{0}) =
p^1(A^1|S^1)$; it is assumed that the distributions over actions are 
known and contribute no
information about $\theta$ to the likelihood (recall that this is an
online estimation problem so that action choice is under the control
of the decision maker).  We also assume that $f^1$ is known; in
practice, one might choose to set $f^1$ to be a point mass at the
observed first state.

Let $\widehat{\theta}_{t}$ denote the maximum likelihood estimator of
$\theta^*$ based on data accumulated during the first $t$ time points.
We assume that $\sqrt{t}(\widehat{\theta}^{t} - \theta^*)$ is
asymptotically normal with mean zero and asymptotic variance-covariance 
$\Omega(\theta^*)$; conditions under which this holds for
data that are not independent and identically distributed have been
studied extensively \citep[][]{silvey1961note,
  billingsley1960statistical, bar1971asymptotic, crowder1976maximum,
  heijmans1986consistent}. Furthermore assume that there exists
$\widehat{\Omega}^{t}$ which converges in probability to 
$\Omega(\theta^*)$.  Under these assumptions, an approximate
Thompson sampling algorithm can be constructed as follows.

For any $\theta\in\Theta$ let $\mathbb{E}_{\theta}$ denote expectation
under parameter value $\theta$ and for any $T > 1$ and $\pi\in\Pi$
define
$V_{\theta}^{T}(\pi) =
\mathbb{E}_{\theta}\sum_{t=1}^{T}\gamma^{t-1}U^{*t}(\pi)$.
Using (\ref{integralEqn}), $V_{\theta}^{T}(\pi)$ can be computed to
arbitrary precision using Monte Carlo methods.  Let
$\widetilde{\theta}^1$ denote a starting value which might be drawn
from a prior distribution, estimated from historical data, or elicited
from domain experts.  Let $\left\lbrace r^t\right\rbrace_{t\ge 1}$ denote
a non-decreasing sequence of positive integers.  
Let $\mathcal{P}_{\Theta}$ denote the orthogonal projection
onto $\Theta$.
Approximate Thompson sampling consists of the
following steps for each time $t$: (i) compute
$\widehat{\pi}^{t} =
\arg\max_{\pi\in\Pi}V_{\widetilde{\theta}^{t}}^{r^t}(\pi)$;
(ii) set $A^t = \widehat{\pi}^t(S^t)$; (iii) observe $S^{t+1}$; 
(iv) draw
$\widecheck{\theta}^{t+1} \sim
\mathrm{Normal}\left\lbrace \widehat{\theta}^{t+1},
  {(t+1)}^{-1}\widehat{\Omega}^{t+1}\right\rbrace$; and (v)
set $\widetilde{\theta}^{t+1} = \mathcal{P}_{\Theta}\widecheck{\theta}^{t+1}$.  

The foregoing algorithm is called `approximate' Thompson sampling
because at each $t$: (i) calculation of the discounted mean utility is truncated
at $r^t$; and (ii) the estimated sampling distribution of the maximum likelihood
estimator is used as an approximation of the posterior distribution of
$\theta$.  A fully Bayesian implementation of Thompson sampling, with
truncation, would draw a strategy $\widetilde{\pi}^t$ from the
posterior distribution of $\arg\max_{\pi\in\Pi}V_{\theta}^{r^t}(\pi)$ at
each time $t$; in settings where evaluation of the likelihood
is expensive, these approximations can result in considerable computational 
savings.  In the next section, we show that approximate Thompson sampling
is consistent and characterize how the rate of convergence depends
on the sequence $\left\lbrace r^t\right\rbrace_{t\ge 1}$.  

\begin{remark}
While we focus on maximum likelihood estimation, the results in the
next section apply directly to any estimator that is asymptotically
normal and satisfies the assumed regularity conditions.   It is possible to obtain even
more generality by assuming that there exists a sequence of positive 
constants $\left\lbrace
  \alpha^t\right\rbrace_{t\ge 1}$ such that
$\alpha^{t}||\widehat{\theta}^{t} - \theta^*||_2 = O_P(1)$ and modifying
the approximate Thompson sampling algorithm to draw samples
$\widetilde{\theta}^{t+1} = \widehat{\theta}^{t+1} +
\mathrm{U}^t/\alpha^{t}$ where $\left\lbrace U^t\right\rbrace_{t\ge
  1}$ are independently and identically distributed sub-gaussian
random variables with mean zero.   However, for simplicity we focus our developments
on the case of $\sqrt{t}$-consistent, asymptotically normal
estimators.  
\end{remark}

\section{Asymptotic properties}  
In this section, we derive rates of convergence for the regret
$V_{\theta^*}(\pi^{\mathrm{opt}}) - V_{\theta^*}(\widehat{\pi}^t)$.
Define $H^t = (\overline{S}^t, \overline{A}^t)$ to be the information
accumulated at time $t$.  In addition to (A0)-(A3), we assume that for
all $t\ge 1$: (A4)  $|U^t| \le 1$ with probability one; (A5) the
parameter space $\Theta = \mathbb{R}^{q}$; (A6)
$\mathbb{E}t^{1/2}(\widetilde{\theta}^t - \theta^*) = O(1)$; and (A7) for $\forall \theta\in \Theta$, the log likelihood
$\ell^{t}(\theta) = \log\mathcal{L}^t(\theta)$ satisfies $|\ell^t(\theta)-\ell^t(\theta^*)|\leq |(\theta-\theta^*)^TD^t(\theta^*)|$, where $\mathbb{E}t^{-1/2}||D^t(\theta^*)||_2=O(1)$.
These assumptions are mild;
 (A6)
holds if $\widehat{\theta}^t$ is regular and asymptotically normal so
that the approximate Thompson sampling algorithm is drawing samples
from a $t^{-1/2}$-neighborhood of $\theta^*$; and (A7) holds under
smoothness and moment conditions on the likelihood
\citep[][]{heijmans1986first}. 
Assumption (A5)
simplifies our arguments but is not necessary; e.g., one could
take $\Theta$ to be a compact subset of $\mathbb{R}^q$ and have
$\theta^*$ be an interior point of $\Theta$.  
 The following result
is proved in the Supplemental Material.
\begin{theorem}\label{theoremey}
Assume (A0)-(A7) and let $\left\lbrace r^t\right\rbrace_{t \ge 1}$ be
a sequence of non-decreasing positive integers.  Then,
\begin{equation*}
V_{\theta^*}(\pi^{\mathrm{opt}}) - V_{\theta^*}(\widehat{\pi}^t) =
O_P\left\lbrace
\left(\frac{r^t}{t}\right)^{1/2} + \gamma^{r^t}
\right\rbrace.
\end{equation*}
Thus, if $r^t = \lfloor -\log(t)/\log(\gamma^2)\rfloor$ then the right-hand-side
is  $O_P\left[\left\lbrace
    \log(t)/t\right\rbrace^{1/2}\right]$.  
\end{theorem}\noindent
The preceding result illustrates the bias-variance trade-off
associated
with choosing $\left\lbrace r^t\right\rbrace_{t \ge 1}$;  as 
$r^t$ increases, the Monte Carlo error incurred by
approximating $V_{\widehat{\theta}^t}$ with
$V_{\widehat{\theta}^t}^{r^t}$ decreases, however, as $r^t$ increases,
the error in approximating $V_{\theta^*}^{r^t}$ with
$V_{\widehat{\theta}^t}^{r^t}$ also increases.  Setting
$r^t = \lfloor -\log(t)/\log(\gamma^2)\rfloor$ optimally balances this trade-off in
that it leads to the fastest rate of convergence.

The preceding result can be sharpened under additional assumptions
on the behavior of $t^{1/2}(\widetilde{\theta}^t -\theta^*)$.  
For any $\theta \in \Theta$, define $\pi_{\theta}^{\mathrm{opt}} = 
\arg\max_{\pi\in\Pi}V_{\theta}(\pi)$ so that $\pi^{\mathrm{opt}} = \pi_{\theta^*}^{\mathrm{opt}}$.
In addition, for any $\epsilon  > 0$ define
\begin{equation*}
R_{\theta^*}(\epsilon) = \sup_{\theta\,:\,\mid\mid
  \theta-\theta^*\mid\mid \le \epsilon}
\left\lbrace
V_{\theta^*}(\pi_{\theta^*}^{\mathrm{opt}}) - V_{\theta^*}(\pi_{\theta}^{\mathrm{opt}})
\right\rbrace,
\end{equation*}  
so that $R_{\theta^*}(\epsilon)$ measures the worst-case regret in an
$\epsilon$-neighborhood of $\theta^*$.  For any $\delta \ge 0$, define
the radius of regret as
$R_{\theta^*}^{-}(\delta) = \sup\left\lbrace \tau \ge 0 \,:\,
  R_{\theta^*}(\tau) \le \delta \right\rbrace$.  Thus,
the radius $R_{\theta^*}^{-}(\delta)$ measures how close $\theta$ must 
be to $\theta^*$ to ensure that $V_{\theta^*}(\pi^{\mathrm{opt}}) - 
V_{\theta^*}(\pi_{\theta}^{\mathrm{opt}})$ is no more than $\delta$,
i.e., problems with a smaller radius are harder in the sense that a more
accurate estimator of $\theta^*$ is required to ensure the same 
performance.  This notion is formalized in the 
 results that follow.  

We strengthen (A6) to (A6') 
${t}^{1/2}(\widetilde{\theta}^{t}-\theta^*) =
t^{-1/2}\sum_{v=1}^{t}\phi^{v}(H^v; \theta^*) + W^t$,
where $\underset{v\in\mathcal{T}}{\sup}||\phi^{v}(H^v;\theta^*)||_2$ is bounded with
probability one,
$E\left\lbrace \phi^{v}(H^v;\theta^*)\mid H^{v-1}\right\rbrace = 0$
with
probability one,  
and there exists $\varsigma, \sigma > 0$ such that
$\mathbb{E}\exp\left(\lambda t^{\varsigma}W_j^{t}\right) 
\le \exp(\lambda^2\sigma^2/2),\,j=1,\ldots, q$, 
for all $t$ and $\lambda \in\mathbb{R}$. We modify Assumption (A7) to (A7') $\ell^t(\theta) - \ell^t(\theta^*) =
(\theta-\theta^*)^{\T}D^t(\theta^*)
+ Q^{t}(\theta, \theta^*)$ where 
$\underset{t\in \mathcal{T}}{\sup}||D^t(\theta^*)/\sqrt{t}||_2$ is bounded with probability one 
and $|Q^t(\theta, \theta^*)| \le tM||\theta-\theta^*||_2^{1+\eta}$ for
some $\eta >0$, $M>0$ and all $t\in \mathcal{T}$.
Proofs of the following results are in the Supplemental Material.
\begin{theorem}\label{expBound}
Assume (A0)-(A5), (A6') and (A7') then for any $\delta > 0$
\begin{multline*}
\mathrm{pr}\left\lbrace
V_{\theta^*}(\pi^{\mathrm{opt}}) - V_{\theta^*}(\widehat{\pi}^{t}) >
\delta
\right\rbrace \le K_1 \exp\left[
  -t \left\lbrace R_{\theta^*}^{-}(\delta/5)\right\rbrace^2 K_2
\right] \\ + 
K_1\exp\left[
-\frac{t}{(r^t)^2}\left\lbrace
(1-\gamma^{r^t})\left[
\frac{K_3}{\sqrt{r^t}} + K_4\left\lbrace 
R_{\theta^*}^{-}(\delta/5)
\right\rbrace ^{\eta}
\right]
\right\rbrace^{-2}\delta^2K_5
\right]
,
\end{multline*}
provided $r^t \ge \log\left(
2\delta(1-\gamma)/15
\right)/\log\gamma$, 
where $K_1, K_2, \ldots,K_5$ are constants that depend on the
dimension $p$ and the discount factor $\gamma$ but not on $\delta$ or
$t$.   
\end{theorem}
\begin{corollary}
Assume (A0)-(A5) and (A6'-A7').  Furthermore, suppose that
$R_{\theta^*}^{-}(0) > 0$ and 
define $\Pi_{\theta^*}^{\mathrm{opt}} = \left\lbrace \pi\in\Pi\,:\,
  V_{\theta^*}(\pi) = V_{\theta^*}(\pi^{\mathrm{opt}})\right\rbrace$.
Then $pr\left\lbrace
\limsup_{t\rightarrow \infty} \widehat{\pi}^t \notin \Pi_{\theta^*}^{\mathrm{opt}}
\right\rbrace = 0$, i.e., approximate Thompson sampling selects an
optimal decision strategy eventually always with probability one.  
\end{corollary}\noindent
The preceding theorem provides a probability bound on the regret of
approximate Thompson sampling; the dependence on
$R_{\theta}^{-}(\delta)$ is intuitive in that the bound becomes looser
as the radius decreases.   The form of the constants $K_{1},\ldots, K_{5}$ 
are given in the Supplementary Material.  


\begin{remark}
In some settings it may be of interest to consider classes of models
for the conditional distribution of $S^t$ given $S^{t-1}$ and
$A^{t-1}$ in which the model complexity increases with $t$.  Theorem
\ref{theoremey} can be extended  to handle this case provided
the likelihood is sufficiently smooth and a rate of convergence is
available for parameter estimators.  Suppose that 
$f(s^{t+1}\mid s^t, a^t) = f(s^{t+1}\mid s^t, a^t; \theta^*)$ where 
$\theta^* \in \Theta\subseteq \mathbb{R}^{\infty}$ and that for each $k \ge 1$ 
one postulates a class of conditional densities $f_k(s^{t+1}\mid s^t,
a^t;\theta_k)$ indexed by $\theta_k \in\Theta_k\subseteq
\mathbb{R}^k$. Let $\theta_{k}^*$ denote the projection of $\theta^*$
onto $\Theta_k$, let  $\widehat{\theta}_{k}^{t}$ denote the maximum
likelihood estimator of $\theta_{k}^*$, and assume that
$(\widehat{\theta}_{k}^{t} - \theta_{k}^*) = O_P(k^{\beta}/t^{1/2})$
for some $\beta > 0$ \citep[e.g.,][]{newey1997convergence}.  
Furthermore, let $\widetilde{\theta}_{k}^t$
denote random perturbation of $\widehat{\theta}_{k}^{t}$ so that
$(\widetilde{\theta}_{k}^{t} - \theta_{k}^*) =
O_P(k^{\beta}/t^{1/2})$, i.e., $\widetilde{\theta}_{k}^{t} = \mathcal{P}_{\Theta_k}\left\lbrace
\widehat{\theta}_{k}^t + \tau^t k^{\beta}Z_k^t/t^{1/2}\right\rbrace$
where
$\mathcal{P}_{\Theta_k}$ is the orthogonal projection onto $\Theta_k$,
$Z_k^t$ is a standard normal vector, and $\tau^t$ is an
estimator
of the square root of the asymptotic variance-covariance of $\widehat{\theta}^t$.  
Furthermore, define $\ell_{k}^t(\theta_k) = 
\sum_{v=2}^{t}\log\, f_k\left(
S^{t}\mid S^{t-1}, A^{t-1}; \theta_k
\right)$ to be the log-likehood and write $\ell^t(\theta) =
\ell_{\infty}^t(\theta_{\infty})$.    Suppose that for all
$\theta_k\in\Theta_k$, 
$\ell_{k}^t(\theta_k) - \ell_{k}^t(\theta^*) =
(\theta-\theta^*)^{\T}D_k^t(\theta^*) 
+ o(\mid\mid\theta_k-\theta^*\mid\mid)$, where
$D_{k}^t(\theta^*) = O_p(t^{1/2}/k^{\beta})$; and 
$\ell_{k}^t(\theta_k^*)
- \ell^{t}(\theta^*) = O_p(t^{-1/2}k^{-\omega})$ for some $\omega >
0$.  
Then, under mild regularity conditions
\begin{equation*}
V_{\theta^*}(\pi^{\mathrm{opt}}) - V_{\theta^*}(\widehat{\pi}^t) = 
O_p\left[
\frac{(r^t)^{1/2}+ (k^{r^t})^{\beta} }{t^{1/2}}
+ \frac{1}{
(k^{r^t})^{\omega}(r^t)^{1/2}}
+ \gamma^{r^t}
\right].
\end{equation*}
\end{remark}

\section{Illustrative simulation experiments}
We illustrate the finite sample performance of the proposed
approximate Thompson sampling algorithm using two simulated
examples: (i) resource allocation for control of the spread of
influenza using an agent-based compartmental model; and (ii) adaptive
management of mallard populations in the U.S.  The first example
was chosen to illustrate the use of approximate Thompson sampling
in a setting where the set of possible decisions is too large to apply
existing methodologies.  The second example was chosen to examine 
the performance of approximate Thompson sampling in a setting
where classic approximate dynamic programming algorithms can
be applied as a gold standard. 
\subsection{Control of influenza}
In our first illustrative example, we consider a simple agent-based
compartmental model for the daily spread of the flu within a closed
population.  We assume that the population is constant, i.e., there
are no births or deaths.  Each day, every member of the population is
in one of three states: uninfected and susceptible; infected and
contagious; or recovered and neither susceptible nor contagious
\citep[see][and references therein]{keeling2008modeling,
  hens2012modeling}.  We assume that disease transmission can occur
only when an infected and susceptible individual come into contact.
To model the contact process, we assume that individuals can only
make contact through direct links in a social network and that they progress
through different social networks over time according to the day of
week and their current health status.  We assume a large
population-level social network which we term the public network.
Each member in the population is also a member of a fully connected
`family' social network. The set of families is obtained by randomly
partitioning the entire population into of groups of size 1-15;
details for computing such a random partition are in the Supplemental
Material.  Each individual in the population is classified as being a
student, employed, or retired (in this utopia there is no unemployment
and students do not work).  Students are randomly assigned to one of
$n_s$ schools and transition among their family network, school
network, and the public network according to the decision rule in the
left panel of Figure (\ref{studentEmployeeRules}).  Employees are
randomly assigned to one of $n_e$ employers and transition among their
family network, work network, and public network according to the
decision rule in the right panel of Figure
(\ref{studentEmployeeRules}).  Retired persons attend the public space
each day unless they are infected in which case they attend the public
space with probability $p_{r,1}$ and stay home with probability
$1-p_{r,1}$.  Individuals are in their family social network whenever
they are at home.  We generate the public, work, and school networks
according to a Barabasi-Albert, Erdos-Renyi, or a Watts-Strogatz
model \citep[][]{newman2010networks}. 
 While our contact model is quite simple, it serves to
illustrate how the proposed method can be applied to more complex
agent-based systems.

Let $S_{\ell}^{t}$ denote the state of individual
$\ell \in \mathcal{L}= \left\lbrace 1,\ldots, L\right\rbrace$ at time
$t \in\mathcal{T}$.  In our influenza model,
$S_{\ell}^t = (S_{\ell,1}^{t}, S_{\ell, 2}^{t}, S_{\ell,3}^{t})$,
where: $S_{\ell,1}^{t} \in \left\lbrace 0, 1, 2\right\rbrace$ denotes
infection status of individual $\ell$ at time $t$ so that $0$ codes
susceptable, 1 codes infected, and  2 codes recovered;
$S_{\ell,2}^{t} \in \left\lbrace 1,\ldots, 100\right\rbrace$ denotes
the age of subject $\ell$ which is assumed to remain constant for all
$t$; and $S_{\ell, 3}^{t}\in\mathbb{R}$ denotes a measure of 
susceptibility.  We draw the ages of individuals labeled as
being in school independently from a uniform distribution on
$\left\lbrace 0,1,\ldots, 25\right\rbrace$, ages of individuals
labeled as employed are drawn independently from a uniform
distribution on $\left\lbrace 15, 16, \ldots, 65\right\rbrace$, and
the ages of those labeled as retired are drawn independently from a
uniform distribution on $\left\lbrace 50, 51,\ldots 90\right\rbrace$. 
Initial susceptibility for subject $\ell$, $S_{\ell, 3}^{1}$, is drawn
according to the linear model $S_{\ell,3}^{1} = \zeta_0 +
\zeta_1S_{\ell,2}^{2} + \epsilon_{\ell}^{1}$ where the errors 
$\left\lbrace \epsilon_{\ell}^t\right\rbrace_{\ell\in\mathcal{L}}$ are
standard normal random variables that are independent
across subjects and the coefficients $\zeta_0$ and $\zeta_1$ are chosen
so that the initial susceptibility has (unconditional) mean zero and
variance $10$.  Evolution of susceptibility follows a first-order
autoregressive model, $S_{\ell,3}^{t+1} = \rho S_{\ell,3}^{t} + 
\nu A_{\ell}^{t} + \epsilon_{\ell}^t$, where $A_{\ell}^t \in \left\lbrace
0, 1\right\rbrace$ is an indicator that individual $\ell$ received
treatment at time $t$ and $\left\lbrace
  \epsilon_{\ell}^{t}\right\rbrace_{\ell\in\mathcal{L},
  t\in\mathcal{T}}$ are independent normal random variables with mean
zero and variance $0.25$.

Let $\mathcal{I}_{\ell}^t$ denote the set of infected individuals with
whom individual $\ell$ could potentially make contact with at time $t$,
i.e., they are in the same social network at time $t$ and there is
an edge between them.  We assume that the probability that individual
$\ell$ becomes infected at time $t$ is 
\begin{equation*}
1 - \prod_{i \in \mathcal{I}_{\ell}^t}\left\lbrace
1 - p_{c}\mathrm{expit}\left(
\vartheta_0 + \vartheta_1 S_{\ell,2}^{t} + \vartheta_{2}
S_{\ell,3}^{t}
+ \vartheta_{3}A_{\ell}^{t} + \vartheta_{4}A_{i}^t +
\vartheta_{5}A_{i}^tA_{\ell}^t
+ \vartheta_{6}(S_{\ell,2}^t - S_{i,2}^t)^2
\right)
\right\rbrace,
\end{equation*}
where $p_c$ denotes the probability of contact,
$\mathrm{expit}(u) = \exp(u)/\left\lbrace 1+\exp(u)\right\rbrace$, and
$\vartheta_0,, \vartheta_1,\ldots, \vartheta_6$ are unknown
parameters.  Thus, the system dynamics model is indexed by
$\theta = (\zeta_0, \zeta_1, p_{r,1}, p_{s,1}, p_{s,2}, p_{e,1},
p_{e,2}, \rho, \nu, p_{c}, \vartheta_0,\vartheta_1,\ldots,
\vartheta_6)^{\T}$.
Let $\theta^*$ denote the true parameter values, in our simulation
settings we set 
$\theta^* = 
(4.16, -0.119, 
0.5, 0.5, 0.5, 0.5,
0.5, 0.8, -0.01, 0.8, -0.5, \allowbreak-0.01, 0.8, -3.5, -3.5, -6, -0.001)^{\T}
$; 
simulations
with other parameter settings were qualitatively similar and therefore
omitted.  
We implemented approximate Thompson sampling using maximum 
likelihood to estimate $\theta^*$ and the observed Fisher
information to approximate the covariance of 
this estimator.

 Let 
$S^t = \left\lbrace S_{\ell}^t\right\rbrace_{\ell\in\mathcal{L}}$, we
optimized
over a parametric class of policies of the form
\begin{equation}\label{fluClassPolicies}
\left\lbrace \pi(s^t; M, \overline{S}^t, \overline{A}^{t-1}, \eta)
\right\rbrace_{\ell} =\left\lbrace
\begin{array}{cl}
1 &\mbox{ if } \sum_{w \in \mathcal{L}}
1_{\phi_{\ell}^{\T}(s^t; \overline{S}^t, \overline{A}^{t-1})\eta \ge 
\phi_{w}^{\T} (s^t;\overline{S}^t, \overline{A}^{t-1})\eta} > L-M \\
0 & \mbox{ otherwise}, 
\end{array}
\right.
\end{equation}
where $\phi_{\ell}(s^t; \overline{S}^t, \overline{A}^{t-1})\in\mathbb{R}^d$ 
is a, possibly data-dependent, feature vector for individual
$\ell\in\mathcal{L}$ constructed from $s^t$ and 
$\eta\in\mathbb{R}^{d}$.  
Thus, the preceding policy ranks the individuals
according to the score $\phi_{\ell}^{\T}(s^t; \overline{S}^t,
\overline{A}^{t-1})\eta$ 
and then assigns
treatment to the $M$ individuals with largest scores; ties are 
broken randomly.   In this application, the feature 
vector for individual $\ell$ is  $\phi_{\ell}(s^t;\overline{S}^t,
\overline{A}^{t-1}) = 
\left\lbrace 
\phi_{\ell,1}(s^t), 
\phi_{\ell,2}(s^t),
\phi_{\ell,3}(s^t),
\phi_{\ell,4}(s^t),
\phi_{\ell,5}(s^t; \overline{S}^t, \overline{A}^{t-1})
\right\rbrace$, where: $\phi_{\ell, 1}(s^t)\in\left\lbrace 0,
  1\right\rbrace$ 
is an indicator of infection; $\phi_{\ell,2}(s^t)\in \left\lbrace
  0,1\right\rbrace$
is an indicator of susceptibility;  
$\phi_{\ell,3}(s^t) \in \left\lbrace 1,\ldots,
  100\right\rbrace$  is age in years; $\phi_{\ell, 4}(s^t) \in
\mathbb{R}$ is susceptability;  and $\phi_{\ell,5}(s^t;\overline{S}^t,
\overline{A}^{t-1}) =
1_{\phi_{\ell,1}(s^t)=0}\sum_{w\ne\ell}1_{\phi_{w,1}(s^t)=1}\widehat{\delta}_{w,\ell}^t
+
1_{\phi_{\ell,1}(s^t)=1}\sum_{w\ne\ell}1_{\phi_{w,1}(s^t)=0}\widehat{\delta}_{w,\ell}^t$ 
and $\delta_{w, \ell}$ the estimated probability of a contact between
individuals $\ell$ and $w$ at the next time point.  

We measure the performace of Thompson sampling in terms of proportion
of individuals who are infected $T=10$ and $T=20$ days after disease
outbreak.  We select 10\% of the population at random to start as
infected at $t=1$ when disease management begins.  We assume that at
most 20\% of the population can be treated at each time point, e.g.,
$M=\lfloor 0.2L\rfloor$ in the class of policies given in
(\ref{fluClassPolicies}).  To form a baseline for evaluating the
proposed algorithm, we also evaluate the performance
of: (i) no treatment; and (ii) a myopic policy wherein treatment is
applied to the 20\% of the population having the highest estimated
probability of becoming infected at the next time point.  
Table \ref{fluTable} shows the results based on 1000 Monte
Carlo replications.  Thompson sampling resulted in 
significantly fewer infections at the end of the observation
period than no treatment or treating myopically; the advantage
of Thompson sampling appears to increase with population
size.

\begin{figure}
\begin{minipage}{0.45\linewidth}
\begin{align*}
& \mbox{\textbf{If} weekday and not infected \textbf{then:}}\\ &\hspace{0.25in}\mbox{attend school;} \\
& \mbox{\textbf{If} weekday and infected \textbf{then:}}\\ 
&\hspace{0.25in}\mbox{attend school with probability $p_{s,1}$ and}\\ &
  \hspace{0.25in}                                                             
  \mbox{stay home with probability $1-p_{s,1}$;}\\
& \mbox{\textbf{If} weekend and not infected \textbf{then:}}\\ &
\hspace{0.25in}\mbox{attend public space;
}\\ 
& \mbox{\textbf{If} weekend and infected \textbf{then:}}\\ &
\hspace{0.25in}\mbox{attend public space with probability $p_{s,2}$ and}\\ &
\hspace{0.25in}
\mbox{stay home with probability $1-p_{s,2}.$
}
\end{align*}
\end{minipage}
\hspace{0.05\linewidth}
\begin{minipage}{0.45\linewidth}
\begin{align*}
& \mbox{\textbf{If} weekday and not infected \textbf{then:}}\\ &\hspace{0.25in}\mbox{attend work;} \\
& \mbox{\textbf{If} weekday and infected \textbf{then:}}\\ 
&\hspace{0.25in}\mbox{attend work with probability $p_{e,1}$ and}\\ &
  \hspace{0.25in}                                                             
  \mbox{stay home with probability $1-p_{e,1}$;}\\
& \mbox{\textbf{If} weekend and not infected \textbf{then:}}\\ &
\hspace{0.25in}\mbox{attend public space;
}\\ 
& \mbox{\textbf{If} weekend and infected \textbf{then:}}\\ &
\hspace{0.25in}\mbox{attend public space with probability $p_{e,2}$ and}\\ &
\hspace{0.25in}
\mbox{stay home with probability $1-p_{e,2}.$
}
\end{align*}
\end{minipage}
\caption{\label{studentEmployeeRules}\textbf{Left:} 
decision rule for a student in agent-based influenza model.
\textbf{Right:} decision rule  for an employed person in the agent-based
influenza model.
}
\end{figure}

\begin{table}
  \def~{\hphantom{0}}
  {\caption{\label{fluTable}Proportion of infected individuals $T$
      days post outbreak.  Results are reported under: 
      (i) no treatment; (ii) assigning treatment to the 20\% of population
      with highest estimated probability of infection at the next time 
      point (Myopic); 
      and  (iii) assigning treatment to 20\% of the population 
      using approximate Thompson sampling (Thompson).   
      The work, school, and public network types are generated
      using a Barabasi-Albert (BA), Erdos-Renyi (ER), or
      Watts-Strogatz (WS) model.  
      Estimates are based on  1000 Monte Carlo replications.
    }}
  \centering
\begin{tabular}{llccc}
\multicolumn{5}{c}{$T=10$}\\
Network & Popn. size & No Treatment & Myopic & Thompson 
sampling  \vspace{0.1in} \\ 
BA & 100 & 0.402 (0.000451) & 0.272 (0.000279) & 0.175 (0.000913) \\
BA & 1000 & 0.452 (0.000413) & 0.311 (0.000395) & 0.199 (0.000553) \\
BA & 10000 & 0.463 (0.005632) & 0.334 (0.004502) & 0.216 (0.007015) \\
BA & 100000 & 0.470 (0.000336) & 0.343 (0.000517) & 
0.231 (0.004364)  \vspace{0.1in} \\ 
WS & 100 & 0.409 (0.002713) & 0.289 (0.000308) & 0.222 (0.000449) \\
WS & 1000 & 0.471 (0.007324) & 0.346 (0.004761) & 0.261 (0.008027) \\
WS & 10000 & 0.479 (0.000531) & 0.377 (0.006054) & 0.283 (0.003962) \\
WS & 100000 & 0.492 (0.005904) & 0.387 (0.006382) 
& 0.291 (0.005043) \vspace{0.1in} \\
ER & 100 & 0.401 (0.000928) & 0.302 (0.000969) & 0.194 (0.006406) \\
ER & 1000 & 0.446 (0.000826) & 0.353 (0.005216) & 0.247 (0.005913) \\
ER & 10000 & 0.454 (0.000574) & 0.365 (0.003182) & 0.258 (0.004216) \\
ER & 100000 & 0.465 (0.000776) & 0.382 (0.004277) & 
0.273 (0.005328)  \vspace{0.10in} \\ 
\multicolumn{5}{c}{$T=20$}\\
Network & Popn. size & No Treatment & Myopic & Thompson 
sampling  \vspace{0.1in} \\ 
BA & 100 & 0.340 (0.000104) & 0.187 (0.003379) & 0.0895 (0.000739) \\
BA & 1000 & 0.364 (0.000375) & 0.201 (0.000603) & 0.113 (0.000706) \\
BA & 10000 & 0.375 (0.000425) & 0.224 (0.000954) & 0.121 (0.001472) \\
BA & 100000 & 0.382 (0.000316) & 0.232 (0.000541) & 
0.140 (0.003049) \vspace{0.1in} \\
WS & 100 & 0.326 (0.000523) & 0.191 (0.000237) & 0.109 (0.000837) \\
WS & 1000 & 0.368 (0.000416) & 0.203 (0.001824) & 0.116 (0.000635) \\
WS & 10000 & 0.372 (0.000865) & 0.220 (0.002083) & 0.132 (0.003346) \\
WS & 100000 & 0.385 (0.001206) & 0.228 (0.001724) & 
0.143 (0.002501) \vspace{0.1in} \\
ER & 100 & 0.317 (0.000178) & 0.177 (0.000354) & 0.127 (0.000413) \\
ER & 1000 & 0.341 (0.000773) & 0.239 (0.000536) & 0.152 (0.004086) \\
ER & 10000 & 0.357 (0.000692) & 0.253 (0.000891) & 0.164 (0.003641) \\
ER & 100000 & 0.366 (0.000565) & 0.265 (0.000672) & 
0.181 (0.005103) \vspace{0.1in} \\
\end{tabular}
\end{table}

\subsection{Management of mallard populations in the U.S.}
In our second illustrative example,  we consider adaptive management of
mallards  in the United States.    The United States Fish and Wildlife
Service began an adaptive harvest program in 1995 wherein 
measurements of current species abundance and ecological conditions
are used to inform the allocation of waterfowl hunting licenses 
\citep[][]{johnson2015multilevel}.  
The goal is to maximize the longterm, cumulative harvest.  
During the past two decades, data
collected as part of this program has been used to create and validate
high-quality system dynamics models for mallard populations.   
Current practice is to use these dynamics models with approximate
dynamic programing to select from among four types of 
harvest practices: (i) liberal; (ii) moderate; (iii) restricted; and
(iv) closed.  For each of these harvest practices, different sets of 
guidelines are passed to individual agencies who set specific harvest
limits based on these guidelines \citep[][]{fish2016ahm}.  
Here we examine the performance of using approximate Thompson sampling to pick
among these harvest practices each season.  Our intent is to
demonstrate that approximate Thompson sampling is competitive 
even in small domains where approximate dynamic programming can 
be directly applied.  

The mallard population dynamics model we use here is based on 
the 2016 United States Fish and Wildlife Service model 
\citep[it is simplified in that we consider a single fly-way, see][for additional
details]{fish2016ahm}.   For each $t=1,2,\ldots$, let $N_{t, \mathrm{AM}}, N_{t, \mathrm{AF}},
N_{t,\mathrm{YM}}, \mbox{ and } N_{t, \mathrm{YF}}$ denote the
number of adult male, adult female, young male, and young females in
the population.  Furthermore, for each $t=1, 2,\ldots$, 
define $P_{t}\in\mathbb{N}$ to be the number of ponds,
$R_t\in (0,1)$ to be the reproductive rate, and $A^t\in\left\lbrace
\mbox{liberal, moderate, restricted, closed}
\right\rbrace$ to be  the harvest practice
at time $t$.  Under harvest practice $A^t=a^t$, let 
$\psi_{\mathrm{AM}}(a^t), \psi_{\mathrm{AF}}(a^t), \psi_{\mathrm{YM}}(a^t)$,
and $\psi_{\mathrm{YF}}(a^t)$ be the survival rates 
for adult males, adult
females, young males, and young females respectively; these
survival rates are assumed to be known and are provided in the
Supplemental Material.  
The total population $N_t = N_{t,\mathrm{AM}}+N_{t,
  \mathrm{AF}} + N_{t, \mathrm{YM}} + N_{t, \mathrm{YF}}$ is assumed
to evolve according to the difference equations
\begin{eqnarray*}
N_{t+1, \mathrm{AM}} &=& N_{t,\mathrm{AM}}\psi_{\mathrm{AM}}(A^t) +
                         0.897
                         N_{t,\mathrm{AM}} R_{t}\psi_{\mathrm{YM}}(A^t) \\ 
N_{t+1,\mathrm{AF}} &=& N_{t, \mathrm{AF}}\left\lbrace \psi_{\mathrm{AF}}(A^t) +
                        R_{t}\psi_{\mathrm{YF}}(A^t)
                        \right\rbrace \\ 
N_{t+1, \mathrm{YM}} &=& N_{t, \mathrm{YM}} \psi_{\mathrm{AM}}(A^t) + 0.897
N_{t, \mathrm{YF}}R_t\psi_{\mathrm{YM}}(A^t) \\ 
N_{t+1,\mathrm{YF}} &=& N_{t,\mathrm{YF}}\left\lbrace
                        \psi_{\mathrm{AF}}(A^t)+R_t\psi_{\mathrm{YF}}(A^t)
\right\rbrace \\
P_{t+1} &\sim& \mathrm{Normal}(\beta_0 + \beta_1P_t,  0.25^2) \\ 
R_{t} &=& 0.7166 + 0.1083P_t - 0.0373N_t, \\
\end{eqnarray*}
where $\beta_0, \beta_1 \in\mathbb{R}$ are unknown parameters.

We compare Thompson sampling fit with maximum likelihood with the
following strategies: (i) always apply a liberal harvesting practice (Liberal);
(ii) always apply a moderate harvesting practice (Moderate); (iii) always apply a 
restricted harvesting practice (Restricted); and (iv) approximate dynamic
programming fit using radial basis functions (Approximate DP).
The details of the proposed approximate dynamic programming procedure
are provided in the Supplemental Material.  In each simulation
setting, we generate historical data from 1995-2016 under the assumed
model with $(\beta_0, \beta_1) =  (2.2127, 0.3420)$  and then simulate
management of the mallard population using each of the foregoing
strategies over the next fifteen years.     We varied the population
size in 1995 from 6.0 million to 13.0 million and assumed that
the management style from 1995-2016 followed actual harvest 
decisions applied by the United States Fish and Wildlife service.

The estimated average cumulative harvest over the fifteen year management period
are displayed in Table \ref{mallardTable}.  These results are based on
1000 Monte Carlo replications.  For each initial population size, approximate
Thompson sampling is competitive with approximate dynamic programming
which is the current gold standard in this domain; dynamic programming
does not scale well and thus cannot be applied to larger problems
like the influenza model.   Both approximate
Thompson sampling and approximate dynamic programming 
perform similarly to the liberal harvest strategy as both recommend
this harvest practice often; frequent recommendation of the
liberal harvest practice is consistent with actual
recommendations of the United States Fish and Wildlife Service over
the past 20 years
\citep[][]{johnson2015multilevel}.  

\begin{table}
\centering
\caption{\label{mallardTable}
Estimated average total harvest during fifteen year management 
period under the following policies: (i) assigning a liberal harvest
practice every year (Liberal); (ii) assigning a moderate harvest
practice every year (Moderate); (iii) assigning a restricted harvest
practice every year (Restricted); (iv) approximate dynamic programming
with radial basis functions (Approximate DP); and (iv) approximate
Thompson Sampling (Thompson sampling).  
Estimates are 
 based on 1000 Monte Carlo replications
}
\begin{tabular}{lccccc}
Initial popn. size & Liberal & Moderate & Restricted &
                                                                Approximate DP 
& Thompson sampling  \vspace{0.1in}\\
6.0 & 11.47 (0.314) & 10.67 (0.214) & 7.31 (0.085) &
 11.48 (0.279) & 11.58 (0.293) \\
7.0 & 12.35 (0.335) & 11.44 (0.226) & 7.76 (0.089) &  12.36 (0.297) & 12.47 (0.313) \\
8.0 & 13.11 (0.352) & 12.11 (0.237) & 8.14 (0.091) &  13.12 (0.313) & 13.23 (0.328) \\
9.0 & 13.80 (0.367) & 12.71 (0.246) & 8.49 (0.094) & 13.81 (0.325) & 13.92 (0.342) \\
10.0 & 14.41 (0.378) & 13.24 (0.253) & 8.79 (0.096) & 14.42 (0.336) & 14.54 (0.354) \\
11.0 & 14.97 (0.388) & 13.72 (0.259) & 9.07 (0.097) & 14.98 (0.345) & 15.10 (0.363) \\
12.0 & 15.47 (0.397) & 14.16 (0.265) & 9.32 (0.099) & 15.49 (0.352) & 15.60 (0.372) \\
13.0 & 15.94 (0.404) & 14.56 (0.270) & 9.54 (0.100) & 15.95 (0.359) & 16.07 (0.379) \\
\end{tabular}
\end{table}

\section{Discussion}
We proposed a variant of Thompson sampling that can be applied to
inform decision making in online contexts with potentially
high-dimensional state or decision spaces.  The proposed algorithm is
simple to implement and can be applied to essentially any sequential
decision problem with a posited class of dynamics models and an
estimator of parameters indexing the underlying dynamics model.
Thus, we believe that Thompson sampling might prove useful
as a general-purpose tool for a wide range of decision problems.  

There are a number of ways in which this work might be extended. We briefly
discuss one such extension that we feel is a pressing open problem.
Thompson sampling maintains positive support across all feasible
decisions, therefore, it may, by random chance, select a decision that
would be viewed as unacceptable by domain experts, e.g., selecting a
closed harvest practice when mallard populations are at a record high.
In such settings, experts may choose to `override' the decision
selected by Thompson sampling thereby disrupting the
exploration-exploitation trade-off used by the algorithm to learn.
Thus, an important extension of Thompson sampling would be one that
could accommodate evolving, possibly data-dependent, constraints on
the available decisions.  Such an extension may have to redefine its
notion of optimality as the policy that leads to maximal marginal mean
outcome may no longer be identifiable under such constraints.  We are
currently pursuing such an extension.

\section*{Acknowledgements}
Eric Laber acknowledges support from 
the National Science Foundation 
(DMS-1555141, DMS-1557733, DMS-1513579)
and the National Institutes of Health 
(1R01 AA023187-01A1, P01 CA142538).

\section*{Supplementary material}

Supplementary material available at \Bka\ online includes proofs for
Theorems 1 and 2 as well as R code to replicate simulation results
presented in Section 4. 

\bibliographystyle{biometrika}
\bibliography{paper-ref}

\end{document}